\documentclass[showpacs,superscriptaddress,prb,twocolumn]{revtex4}
\usepackage{amssymb}
\usepackage{amsmath}
\usepackage{graphicx}

\begin{document}

\title{Massive Dirac fermions and spin physics in an ultrathin film of
topological insulator}
\date{\today}
\author{Hai-Zhou Lu}
\affiliation{Department of Physics and Center for Theoretical and Computational Physics,
The University of Hong Kong, Pokfulam Road, Hong Kong, China}
\author{Wen-Yu Shan}
\affiliation{Department of Physics and Center for Theoretical and Computational Physics,
The University of Hong Kong, Pokfulam Road, Hong Kong, China}
\author{Wang Yao}
\affiliation{Department of Physics and Center for Theoretical and Computational Physics,
The University of Hong Kong, Pokfulam Road, Hong Kong, China}
\author{Qian Niu}
\affiliation{Department of Physics, The University of Texas, Austin, Texas 78712-0264,
USA }
\author{Shun-Qing Shen}
\affiliation{Department of Physics and Center for Theoretical and Computational Physics,
The University of Hong Kong, Pokfulam Road, Hong Kong, China}

\begin{abstract}
We study transport and optical properties of the surface states which lie in
the bulk energy gap of a thin-film topological insulator. When the film
thickness is comparable with the surface state decay length into the bulk,
the tunneling between the top and bottom surfaces opens an energy gap and
form two degenerate massive Dirac hyperbolas. Spin dependent physics emerges
in the surface bands which are vastly different from the bulk behavior.
These include the surface spin Hall effects, spin dependent orbital magnetic
moment, and spin dependent optical transition selection rule which allows
optical spin injection. We show a topological quantum phase transition where
the Chern number of the surface bands changes when varying the thickness of
the thin film.
\end{abstract}

\maketitle



\section{Introduction}

A three-dimensional (3D) topological insulator is a novel quantum state of
matter which possesses metallic surface states in the bulk energy gap.\cite%
{Fu-07prl,Moore-07prb,Murakami-07NJP,Teo-08prb} The surface states consist
of an odd number of massless Dirac cones which are protected by $Z_{2}$
topological invariants. The first 3D topological insulator is Bi$_{x}$Sb$%
_{1-x}$, an alloy with complex structure of surface states which was
confirmed by using angle-resolved photoemission spectroscopy (ARPES).\cite%
{Hsieh-08Nature,Hsieh-09Science,Nishide-09xxx} Recently it was verified that
Bi$_{2}$Se$_{3}$ and Bi$_{2}$Te$_{3}$ have only one Dirac cone near the $%
\Gamma $ point by both experiments and the first principles calculation,\cite%
{Xia-09NP,Zhang-09NP,Hsieh-09Nature,Chen-09Science} which attracts extensive
attentions in condensed matter physics. The electrons or Dirac fermions in
the surface states of topological insulator obey the 2+1 Dirac equations and
reveal a lot of unconventional properties such as the topological
magnetoelectric effect.\cite{Qi-08prb} It was also proposed that the surface
states interfaced with a superconductor can form Majorana fermions for
performing fault-tolerant quantum computation.\cite%
{Fu-08prl,Nilsson-08prl,Fu-09prl,Akhmerov-09prl,Tanaka-09prl,Law-09prl}
Since the surface states surround the sample, it is still a great challenge
for both experimentalists and theorists to explore the transport properties
for metallic surface states in the topological insulators.

In this paper, we study an ultrathin film of topological insulator where
tunneling between the surface states on the top and bottom surfaces opens a
finite gap in the Dirac cone centered at the $\Gamma $ point ($k=0$). The
low lying physics of the ultrathin film can be described as two degenerate
massive Dirac hyperbolas which form time reversal copy of each other. Each
massive band has a $\mathbf{k}$-dependent spin configuration: one near the $%
\Gamma $ point determined by the energy gap, and the other at $k$ large
enough determined by a spin-orbit coupling term quadratic in $k$. We show
that the energy gap oscillates with the thin film thickness, and changes
sign at critical thicknesses. Across the transition points, the $\mathbf{k}$%
-dependent spin configuration near the $\Gamma $ point is flipped while
those at large $k $ remains unchanged, leading to a topological quantum
phase transition where the Chern numbers of the surface bands change [Eq. (%
\ref{chernnumber})]. In the two Dirac hyperbolas of opposite spin
configurations, the $\mathbf{k}$-dependent spin structure results in a
distribution of orbital magnetic moment and Berry curvature with opposite
signs. In doped metallic regime, the Berry curvature drives the spin Hall
effect of the extra carriers which leads to net spin accumulations on the
thin film edges. We also discover a spin-dependent optical transition
selection rule which allows optical injection of spins in the thin film.

The paper is organized as follows. In Sec. \ref{sec:model}, we present the
derivation of the effective Hamiltonian for the thin film of topological
insulator. In Sec. \ref{sec:gapspin}, the oscillation of the gap and the $%
\mathbf{k}$-dependent spin configuration are discussed. In Sec. \ref%
{sec:berry}, the orbital magnetic moment and Berry curvature are addressed.
In Sec. \ref{sec:hall}, the spin Hall conductance is derived in detail. In
Sec. \ref{sec:optical}, the spin-dependent optical transition selection rule
is shown. Finally, a conclusion is given in Sec. \ref{sec:conclusion}.

\section{\label{sec:model}Model for an ultrathin film}

We start with the low-lying effective model for bulk Bi$_{2}$Se$_{3}$, in
which surface states consist of a single Dirac cone at the $\Gamma $ point.%
\cite{Zhang-09NP} We take the periodic boundary conditions in the $x$-$y$
plane such that $k_{x}$ and $k_{y}$ are good quantum numbers, and denote the
thickness of the thin film along $z$ direction as $L$. In the basis of \{$%
\left\vert p1_{z}^{+},\uparrow \right\rangle $,$\left\vert
p2_{z}^{-},\uparrow \right\rangle $,$\left\vert p1_{z}^{+},\downarrow
\right\rangle $,$\left\vert p2_{z}^{-},\downarrow \right\rangle \}$ which
are the hybridized states of Se p orbital and Bi p orbital, with even ($+$)
and odd ($-$) parities, the model Hamiltonian is given by
\begin{equation}
H(k)=(C-D_{1}\partial _{z}^{2}+D_{2}k^{2})+\left(
\begin{array}{cc}
h(A_{1}) & A_{2}k_{-}\sigma _{x} \\
A_{2}k_{+}\sigma _{x} & h(-A_{1})%
\end{array}%
\right)   \label{model}
\end{equation}%
where
\begin{equation*}
h(A_{1})=(M+B_{1}\partial _{z}^{2}-B_{2}k^{2})\sigma _{z}-iA_{1}\partial
_{z}\sigma _{x},
\end{equation*}%
and $\sigma _{\alpha }$ are the Pauli matrices, with $k_{\pm }=k_{x}\pm
ik_{y}$ and $k^{2}=k_{x}^{2}+k_{y}^{2}$. This model is invariant under time
reversal symmetry and inversion symmetry. In this paper, the model
parameters are adopted from Ref. \onlinecite{Zhang-09NP}: $M=0.28$eV$%
,A_{1}=2.2$eV\r{A}, $A_{2}=4.1$eV\r{A}, $B_{1}=10$eV\r{A}$^{2}$, $B_{2}=56.6$%
eV\r{A}$^{2}$, $C=-0.0068$eV, $D_{1}=1.3$eV\r{A}$^{2}$, $D_{2}=19.6$eV\r{A}$%
^{2}$. Though we adopt this concrete model to study the properties of an
ultrathin film of topological insulator, the general conclusion in this
paper should be applicable to other topological insulators.

To establish an effective model for an ultrathin film, we first find the
four solutions to the surface states of the model in Eq. (\ref{model}) at
the $\Gamma $ point ($k_{x}=k_{y}=0$),\cite{Zhou-08prl}
\begin{equation}
H_{0}=\left[
\begin{array}{cc}
h_{0}(A_{1}) & 0 \\
0 & h_{0}(-A_{1})%
\end{array}%
\right] ,  \label{H_0}
\end{equation}%
where
\begin{equation*}
h_{0}(A_{1})=C-D_{1}\partial _{z}^{2}+(M+B_{1}\partial _{z}^{2})\sigma
_{z}-iA_{1}\partial _{z}\sigma _{x},
\end{equation*}%
The solution of the block-diagonal $H_{0}$ can be found by putting a
two-component trial solution into the eigen equation of the upper block
\begin{equation}
h_{0}(A_{1})\left(
\begin{array}{c}
a \\
b%
\end{array}%
\right) e^{\lambda z}=E\left(
\begin{array}{c}
a \\
b%
\end{array}%
\right) e^{\lambda z},
\end{equation}%
with $a$, $b$, $\lambda $ the trial coefficients defining the behavior of
the wavefunctions, and $E$ the trial eigen energy. Note that the trial
coefficients may have multiple solutions, the final solution should be a
linear superposition of these solutions, with the superposition coefficients
determined by boundary conditions. Then the problem becomes a
straightforward calculation of the Schr\"{o}dinger equation. Consider an
open boundary condition that the wavefunctions vanish at the two surfaces
located at $z=\pm L/2$ of the film. Then we finally obtain two
transcendental equations,
\begin{equation}
\frac{\lbrack C-M-E-(D_{1}+B_{1})\lambda _{1}^{2}]\lambda _{2}}{%
[C-M-E-(D_{1}+B_{1})\lambda _{2}^{2}]\lambda _{1}}=\frac{\tanh (\lambda
_{\alpha }L/2)}{\tanh (\lambda _{\bar{\alpha}}L/2)},  \label{trans_eqn}
\end{equation}%
note that where $\alpha $ =1 and 2, $\bar{\alpha}=2$ if $\alpha =1$, vice
versa, so there are two transcendental equations. In Eq. (\ref{trans_eqn}), $%
\lambda _{\alpha }$ define the behavior of the wavefunctions along $z$-axis,
and are functions of the energy $E$
\begin{equation}
\lambda _{\alpha }(E)=\sqrt{\frac{-F+(-1)^{\alpha -1}\sqrt{R}}{%
2(D_{1}^{2}-B_{1}^{2})}},  \label{lambda_alpha}
\end{equation}%
where for convenience we have defined $F=A_{1}^{2}+2D_{1}(E-C)-2B_{1}M$ and $%
R=F^{2}-4(D_{1}^{2}-B_{1}^{2})[(E-C)^{2}-M^{2}]$. The self-consistent
solution to the two equations in (\ref{trans_eqn}) can be found numerically,
and give two energies at the $\Gamma $ point, i.e., $E_{+}$ and $E_{-}$,
which define an energy gap
\begin{equation}
\Delta \equiv E_{+}-E_{-}.  \label{gapdef}
\end{equation}%
Note that the bulk states with much higher or lower energies due to the
quantization of $k_{z}$ in the finite quantum well in principle can also be
obtained in this way, but are ignored here because we concentrate on the
surface states near the gap. The eigen wavefunctions for $E_{+}$ and $E_{-}$
are, respectively,
\begin{eqnarray}
\varphi ^{\uparrow }(A_{1}) &=&C_{+}\left[
\begin{array}{c}
-(D_{1}+B_{1})\eta _{1}^{+}f_{-}^{+} \\
iA_{1}f_{+}^{+}%
\end{array}%
\right] ,  \notag  \label{phichi} \\
\chi ^{\uparrow }(A_{1}) &=&C_{-}\left[
\begin{array}{c}
-(D_{1}+B_{1})\eta _{2}^{-}f_{+}^{-} \\
iA_{1}f_{-}^{-}%
\end{array}%
\right] ,
\end{eqnarray}%
where $C_{\pm }$ are the normalization constants. The superscripts of $%
f_{\pm }^{\pm }$ and $\eta _{1,2}^{\pm }$ stand for $E_{\pm }$, and the
subscripts of $f_{\pm }^{\pm }$ for parity, respectively. The expressions
for $f_{\pm }^{\pm }$ and $\eta _{1,2}^{\pm }$ are given by
\begin{eqnarray}
f_{+}^{\pm }(z) &=&\frac{\cosh (\lambda _{1}^{\pm }z)}{\cosh (\lambda
_{1}^{\pm }L/2)}-\frac{\cosh (\lambda _{2}^{\pm }z)}{\cosh (\lambda
_{2}^{\pm }L/2)},  \notag  \label{feta} \\
f_{-}^{\pm }(z) &=&\frac{\sinh (\lambda _{1}^{\pm }z)}{\sinh (\lambda
_{1}^{\pm }L/2)}-\frac{\sinh (\lambda _{2}^{\pm }z)}{\sinh (\lambda
_{2}^{\pm }L/2)},  \notag \\
\eta _{1}^{\pm } &=&\frac{(\lambda _{1}^{\pm })^{2}-(\lambda _{2}^{\pm })^{2}%
}{\lambda _{1}^{\pm }\coth (\lambda _{1}^{\pm }L/2)-\lambda _{2}^{\pm }\coth
(\lambda _{2}^{\pm }L/2)},  \notag \\
\eta _{2}^{\pm } &=&\frac{(\lambda _{1}^{\pm })^{2}-(\lambda _{2}^{\pm })^{2}%
}{\lambda _{1}\tanh (\lambda _{1}^{\pm }L/2)-\lambda _{2}^{\pm }\tanh
(\lambda _{2}^{\pm }L/2)}.
\end{eqnarray}%
where $\lambda _{\alpha }^{\pm }\equiv \lambda _{\alpha }(E_{\pm })$ can be
found by putting back $E_{\pm }$ into Eq. (\ref{lambda_alpha}).

\begin{figure}[tbph]
\centering  \includegraphics[width=0.46\textwidth]{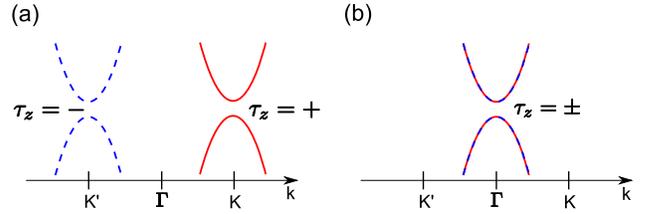}
\caption{(Color online) Schematic comparison between (a) the gapped K-K'
valleys in the staggered graphene, and (b) the two-fold degenerate
hyperbolas described by our effective Hamiltonian in Eq. (\protect\ref{Heff}%
).}
\label{fig:hyperbola}
\end{figure}

By replacing $A_{1}$ by $-A_{1}$ in the above solutions, the energies of the
lower block $h_{0}(-A_{1})$ of $H_{0}$ are found degenerate with those of $%
h_{0}(A_{1})$ and their wave functions are denoted as $\varphi ^{\downarrow
}(-A_{1})$ and $\chi ^{\downarrow }(-A_{1})$, respectively. Now we have four
states, namely, $[\varphi ^{\uparrow }(A_{1}),0]^{T}$, $[\chi ^{\uparrow
}(A_{1}),0]^{T}$, $[0,\varphi ^{\downarrow }(-A_{1})]^{T}$, and $[0,\chi
^{\downarrow }(-A_{1})]^{T}$, where $\uparrow $ and $\downarrow $ imply that
the orbits with spin up and down are decoupled. By using these four
solutions as basis states and rearranging their sequence following (note
that each basis state is a four component vector)
\begin{eqnarray}
&&\left(
\begin{array}{c}
\varphi ^{\uparrow }(A_{1}) \\
0 \\
\end{array}%
\right) ,\left(
\begin{array}{c}
0 \\
\chi ^{\downarrow }(-A_{1}) \\
\end{array}%
\right) ,\left(
\begin{array}{c}
\chi ^{\uparrow }(A_{1}) \\
0 \\
\end{array}%
\right) ,\left(
\begin{array}{c}
0 \\
\varphi ^{\downarrow }(-A_{1}) \\
\end{array}%
\right) ,  \notag \\
&&
\end{eqnarray}%
we can map the original Hamiltonian to the Hilbert space spanned by these
four states, and reach a new low-energy effective Hamiltonian for the
ultrathin film,
\begin{equation}
H_{\mathrm{eff}}=\left[
\begin{array}{cc}
h_{+}(k) & 0 \\
0 & h_{-}(k)%
\end{array}%
\right]   \label{Heff}
\end{equation}%
in which
\begin{eqnarray}
h_{\tau _{z}}(k) &=&E_{0}-Dk^{2}-\hbar v_{F}(k_{x}\sigma _{y}-k_{y}\sigma
_{x})  \notag  \label{htauz} \\
&&+\tau _{z}(\frac{\Delta }{2}-Bk^{2})\sigma _{z}.
\end{eqnarray}%
Note that here the basis states of Pauli matrices stand for spin-up and
spin-down states of real spin. In Eq. (\ref{htauz}), we have introduced a
hyperbola index $\tau _{z}=\pm 1$ (or $\pm $). As shown in Fig. \ref%
{fig:hyperbola}, one can view the hyperbolas as the K and K' valleys in the
staggered graphene [(a)], but being relocated to the $\Gamma $ point[(b)].
Unlike the momentum correspondence in graphene, it is the $\sigma _{z}$ to $%
-\sigma _{z}$ correspondence in the present case. Therefore, the dispersions
of $h_{\pm }$ are actually doubly degenerate, which is secured by
time-reversal symmetry. Here, $\tau _{z}=\pm $ are used to distinguish the
two degenerate hyperbolas, $h_{+}(k)$ and $h_{-}(k)$ describe two sets of
Dirac fermions, each show a pair of conduction and valence bands with the
dispersions
\begin{equation}\label{eigenenergies}
\varepsilon _{c/v}(\mathbf{k})=E_{0}-Dk^{2}\pm \sqrt{(\Delta
/2-Bk^{2})^{2}+(\hbar v_{F})^{2}k^{2}},
\end{equation}
where $c$ and $v$ correspond to the conduction and valence bands,
respectively. The eigen states for $\varepsilon _{c/v}$ are
\begin{equation}\label{eigenstates}
u_{c/v}(\mathbf{k})=\frac{1}{\left\Vert u_{c/v}\right\Vert }\left[
\begin{array}{c}
(\Delta /2-Bk^{2})\tau _{z}+\varepsilon _{c/v} \\
-i\hbar v_{F}k_{+}
\end{array}
\right]
\end{equation}
with $\left\Vert u_{c/v}\right\Vert =\sqrt{[(\Delta /2-Bk^{2})\tau
_{z}+\varepsilon _{c/v}]^{2}+(\hbar v_{F})^{2}k^{2}}.$ Besides the gap $%
\Delta $ already defined in Eq. (\ref{gapdef}), the other parameters in
Hamiltonian (\ref{htauz}) are given by
\begin{eqnarray}
v_{F} &=&(A_{2}/\hbar )\langle \varphi (A_{1})|\sigma _{x}|\chi
(-A_{1})\rangle ,  \notag  \label{paradefinition} \\
D &=&(B_{2}/2)(\langle \varphi ^{\uparrow }|\sigma _{z}|\varphi ^{\uparrow
}\rangle +\langle \chi ^{\uparrow }|\sigma _{z}|\chi ^{\uparrow }\rangle
)-D_{2},  \notag \\
B &=&(B_{2}/2)(\langle \varphi ^{\uparrow }|\sigma _{z}|\varphi ^{\uparrow
}\rangle -\langle \chi ^{\uparrow }|\sigma _{z}|\chi ^{\uparrow }\rangle ),
\notag \\
E_{0} &=&(E_{+}+E_{-})/2,
\end{eqnarray}%
and can be calculated numerically by using Eq. (\ref{phichi}).

\begin{figure}[tbph]
\centering  \includegraphics[width=0.45\textwidth]{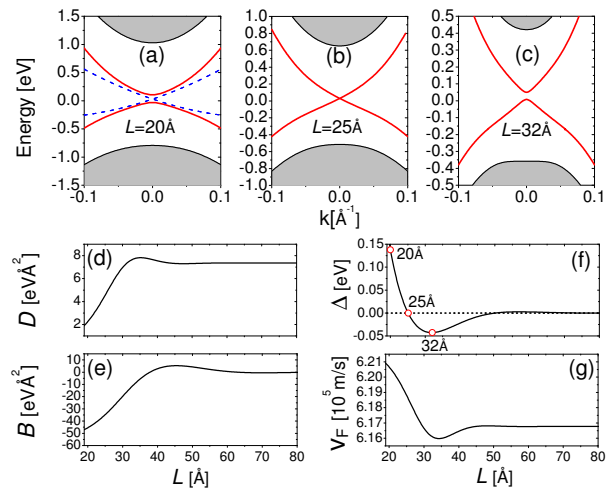}
\caption{(Color online) [(a)-(c)]: Two-fold degenerate ($\protect\tau_z=\pm 1
$) energy spectra of surface states for thickness $L=20$, 25, 32 \AA (solid
lines), and $L\rightarrow \infty $ (dash lines). The grey area corresponds
to the bulk states. The energy spectra are obtained by solving $%
H(k,-i\partial _{z})\Psi (z)=E\Psi (z)$ under the boundary conditions $\Psi
(z=\pm L/2)=0$. Please note that the scales of energy axis in (a)-(c) are
different. The model parameters are adopted from Ref. \onlinecite{Zhang-09NP}%
: $M=0.28$eV$,A_{1}=2.2$eV\AA , $A_{2}=4.1$eV\AA , $B_{1}=10$eV\AA $^{2}$, $%
B_{2}=56.6$eV\AA $^{2}$, $C=-0.0068$eV, $D_{1}=1.3$eV\AA $^{2}$, $D_{2}=19.6$%
eV\AA $^{2}$. [(d)-(g)] The parameters for the new effective model $H_{%
\mathrm{eff}}$: $D$, $B$, the energy gap $\Delta $, and the Fermi velocity $%
v_{F}$ vs $L$. }
\label{fig:gap}
\end{figure}

The numerical results of $\Delta$, $v_{F}$, $D$, and $B$ are presented in
Fig. \ref{fig:gap}. It is noted that $|D|$ must be less than $|B|$,
otherwise the energy gap will disappear, and all discussions in the
following will not be valid. The $\Delta $ terms play a role of mass term in
2+1 Dirac equations.

In the large $L$ limit,
\begin{eqnarray}
v_{F}=(A_{2}/\hbar )\sqrt{1-D_{1}^{2}/B_{1}^{2}}.
\end{eqnarray}
The dispersion relation is given by
\begin{eqnarray}
\varepsilon_{c/v }(k)=\pm v_{F}\hbar k
\end{eqnarray}
for small $k$. As a result, the energy gap closes at $k=0$. The two massless
Dirac cones are located near the top and bottom surfaces, respectively, as
expected in a 3D topological insulator.

In a small $L$ limit,
\begin{eqnarray}
v_{F}=A_{2}/\hbar,
\end{eqnarray}
and
\begin{eqnarray}
\Delta =2B_{1}\pi ^{2}/L^{2}.
\end{eqnarray}
The ratio of the velocity between the two limits is
\begin{eqnarray}
\eta =1/\sqrt{1-D_{1}^{2}/B_{1}^{2}}.
\end{eqnarray}
It is noted that the velocity and energy gap for an ultrathin film are
enhanced for a thinner film.

\section{\label{sec:gapspin}Energy gap and k-dependent spin configuration}

The opening of energy gap for the Dirac fermions is expected as a result of
quantum tunneling between the surface states on the top and bottom surfaces.
When the thickness of the ultrathin film is comparable with the decay length
of the surface states into the bulk, the wavefunctions of the top and bottom
surface states have a spatial overlap which leads to an energy gap at the $%
\Gamma $ point, analogous to the splitting of the bound and anti-bound
orbitals in a double-well potential. The dispersion relation of the surface
states are plotted in Fig. \ref{fig:gap}(a), (b) and (c) for several
thicknesses. A massless dispersion is obtained for a large $L$ limit as
expected. For a finite thickness, the energy gap at $k=0$ is a function of
the thickness $L$ and decays quickly with $L$ [see Fig. \ref{fig:gap}(f)].
It is noticed that the gap $\Delta $ even changes its sign at certain values
of $L$. For instance, for the present case, at about integer times of 25 \AA %
. Correspondingly, the velocity of the Dirac fermions is also thickness
dependent, which is enhanced for a small $L$. Strictly speaking, for an
ultrathin film, these so-called ``surface states" emerge in the entire film.
Nevertheless, they always lie in the bulk energy gap and can thus be
distinguished from those bulk originated quantum well states.

\begin{figure}[tbph]
\centering  \includegraphics[width=0.4\textwidth]{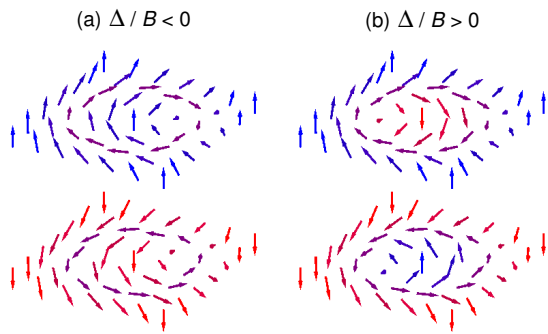}
\caption{(Color online) Schematic illustration of the $\mathbf{k}$-dependent
spin configurations in the conduction band (top) and valence band (bottom)
in Dirac hyperbola $\protect\tau_z=+1$ when (a) $\Delta /B<0$ and (b) $%
\Delta /B>0$. The center and corners of each panel correspond to that $k=0$
and $k$ is large enough, respectively.}
\label{fig:meron}
\end{figure}

From the solution to Eq. (\ref{Heff}), it is obvious that the spin vectors
in each band take a $\mathbf{k}$-dependent spin configuration, in the
neighborhood of the $\Gamma $ point determined by the gap parameter $\Delta $%
, and at large $k$ determined by the term $B\sigma _{z}k^{2}$, as
schematically illustrated in Fig. \ref{fig:meron}. As the two Dirac
hyperbolas are time reversal copy of each other, it is obvious that they
have just the opposite $\mathbf{k}$-dependent spin configurations, as can be
seen from the $\tau_z$ dependence of $h_{\tau_z}(k)$ in Eq. (\ref{htauz}).

\section{\label{sec:berry}Orbital magnetic moment and Berry curvatures}

The opposite $\mathbf{k}$-dependent spin configurations result in opposite
physical properties of the surface Bloch electrons in the two Dirac
hyperbolas, including the orbital magnetic moments and Berry curvatures, as
shown in the context of graphene.\cite{Xiao-07prl} These properties make
possible the manipulation of spin dynamics by electric and magnetic fields
in the thin film topological insulator.

Orbital magnetic moment arises from a self-rotating motion of the surface
Bloch electron and is defined as
\begin{eqnarray}  \label{orbitaldefinition}
m(\mathbf{k})=-i\frac{e}{2\hbar}\left\langle \nabla _{k}u(\mathbf{k}%
)\right\vert \times \lbrack h_{\tau_z}-\varepsilon (k)]\left\vert \nabla
_{k}u(\mathbf{k})\right\rangle \cdot\hat{z} ,
\end{eqnarray}
where $\varepsilon (\mathbf{k})$ and $u(\mathbf{k})$ are the dispersion and
eigenstates of $h_{\tau_z}$.\cite{Chang-08JPCM,Xiao09-arxiv} By putting Eqs.
(\ref{eigenenergies}) and (\ref{eigenstates}) into Eq. (\ref%
{orbitaldefinition}), we find the conduction and valence bands for each $%
h_{\tau_z}$ have the same orbital angular momentum
\begin{equation}
m(k)=-\tau_z\frac{|e|}{\hbar }\frac{\hbar ^{2}v_{F}^{2}(\Delta /2+Bk^{2})}{%
2[(\Delta /2-Bk^{2})^{2}+\hbar ^{2}v_{F}^{2}k^{2}]}.  \label{moment}
\end{equation}%
At the $\Gamma $ point, the two degenerate Dirac hyperbolas acquire opposite
orbital moments which add to the spin magnetic moment.

\begin{figure}[tbph]
\centering \includegraphics[width=0.45\textwidth]{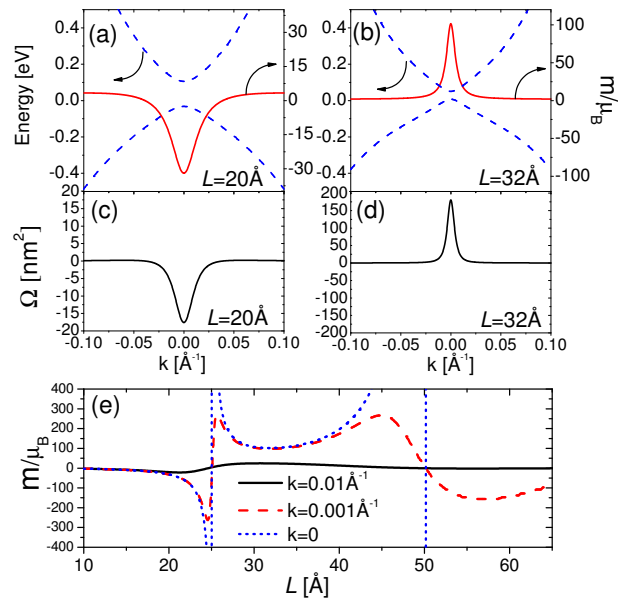}
\caption{(Color online) [(a)(b)]: Energy spectra of surface states (dash
lines) and orbital angular moment $m$ of the conduction bands (solid lines)
as functions of $k$ for $L=20$ and 32 \AA . [(c)(d)] Berry curvature $\Omega
(k)$ of the conduction band. (e) Orbital angular moment $m$ of the
conduction band as a function of thickness $L$ for $k$=0.01, 0.001, and 0 $%
\mathring{A}^{-1}$. The singularities occur at where $\Delta $ changes sign.
Only results for the hyperbola $\protect\tau_z=+1$ are shown. The results
for $\protect\tau_z=-1$ are right opposite to those for $\protect\tau_z=+1$.
}
\label{fig:orbital moment}
\end{figure}

The $\mathbf{k}$-dependent spin configuration also results in a gauge field
in the crystal momentum space, known as the Berry curvature
\begin{eqnarray}
\Omega (\mathbf{k})=\hat{z}\cdot \nabla_{\mathbf{k}}\times \left\langle u(%
\mathbf{k})\right\vert i\nabla _{k}\left\vert u(\mathbf{k})\right\rangle .
\end{eqnarray}
In an in-plane electric field, electrons acquire an anomalous transverse
velocity proportional to the Berry curvature, giving rise to the Hall
effect. In the conduction bands, the Berry curvature distribution near the $%
\Gamma $ point is
\begin{equation}  \label{berrycurvature}
\Omega (k)=-\tau_z\frac{\hbar ^{2}v_{F}^{2}(\Delta /2+Bk^{2})}{2[(\Delta
/2-Bk^{2})^{2}+\hbar ^{2}v_{F}^{2}k^{2}]^{3/2}}.
\end{equation}
The Berry curvature in the valence bands is right opposite to that of in the
conduction bands. There are opposite distributions of the Berry curvature in
the two Dirac hyperbolas. The in-plane electric field can therefore drive
the spin up and spin down electrons towards the opposite transverse edges of
the thin film, which is a surface spin Hall effect.

\section{\label{sec:hall}Spin Hall effect and topological quantum phase
transition}

The $\Delta$ term in our model plays a role as the magnetization in the massive Dirac model exploited to study the anomalous Hall effect.\cite{Sinitsyn2007-prb} In principle, we can find the Hall conductance for each $h_{\tau_z}$. Note that $h_{\tau_z}$ in Eq. (\ref{htauz}) can be explicitly written as
\begin{eqnarray}  \label{hpauli}
h_{\tau_z} &=& E_{0}-Dk^{2}+\sum_{i=x,y,z} d_i \sigma_i,
\end{eqnarray}
where $\sigma_i$ are the Pauli matrices, and the $\mathbf{d}(k)$ vectors
\begin{eqnarray}  \label{dvectors}
d_x &=& \hbar v_F k_y ,\ d_y=-\hbar v_F k_x,\ d_z=\tau_z(\frac{\Delta }{2}%
-Bk^{2}),  \notag \\
\end{eqnarray}

For the $2\times 2$ Hamiltonian in terms of the $\mathbf{d}(k)$ vectors and
Pauli matrices, the Kubo formula for the Hall conductance can be generally
expressed as\cite{Qi2006.prb.74.085308,Zhou2006.prb.73.165303}
\begin{equation}  \label{hallformula}
\sigma _{xy}=\frac{e^{2}}{2 \hbar }\int \frac{d^2\mathbf{k}}{(2\pi)^2}\frac{%
(f_{k,c}-f_{k,v})}{d^{3}}\epsilon _{\alpha \beta \gamma }\frac{\partial d_{a}%
}{\partial k_{x}}\frac{\partial d_{\beta }}{\partial k_{y}}d_{\gamma },
\end{equation}
where $d$ is the norm of $(d_{x},d_{y},d_{z})$, $f_{k,c/v }$ = $1/\{\exp
[(\varepsilon_{c/v}(k)-\mu )/k_{B}T]+1\}$ the Fermi distribution function of
the conduction ($c$) and valence ($v$) bands, with $\mu $ the chemical
potential, $k_{B}$ the Boltzmann constant, and $T$ the temperature.

\begin{figure}[tbph]
\centering \includegraphics[width=0.45\textwidth]{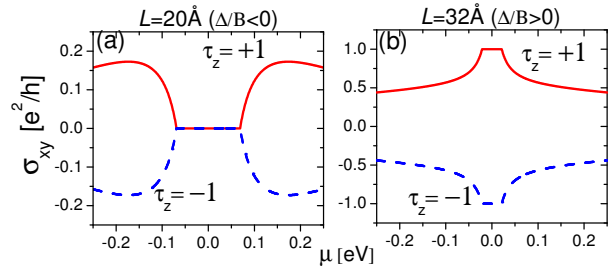}
\caption{(Color online) The hyperbola-dependent Hall conductance vs Fermi
level $\protect\mu$ for $L=20$\AA\ (a) and 32\AA\ (b), respectively.}
\label{fig:hall conductance}
\end{figure}

At zero temperature and when the chemical potential $\mu $ lies between $(-%
\frac{|\Delta |}{2},\frac{|\Delta |}{2})$, the Fermi functions reduce to $%
f_{k,c}=0$ and $f_{k,v}=1$. By substituting Eq. (\ref{dvectors}) into (\ref%
{hallformula}) we arrive at
\begin{eqnarray}  \label{integral1}
\sigma_{xy}(0,\tau_z)=-\tau_z\frac{e^2}{4h}\int_0^{\infty}d(k^2) \frac{%
(\hbar v_F)^2(\frac{\Delta}{2}+Bk^2)}{[(\hbar v_F)^2k^2+(\frac{\Delta}{2}%
-Bk^2)^2]^{\frac{3}{2}}}.  \notag \\
\end{eqnarray}
Note that by comparing above equation with Eq. (\ref{berrycurvature}), we
can write the Hall conductance in the form of the Berry curvature of the
conduction band,
\begin{eqnarray}
\sigma_{xy}(0,\tau_z)=\frac{e^{2}}{\hbar }\int\frac{d^2\mathbf{k}}{(2\pi
)^{2}}(f_{k,v}-f_{k,c})\Omega (k).
\end{eqnarray}
By defining
\begin{equation}
\cos \theta =\frac{(\frac{\Delta }{2}-Bk^{2})}{\sqrt{(\hbar v_{\mathrm{F}%
})^{2}k^{2}+(\frac{\Delta }{2}-Bk^{2})^{2}}},  \label{costheta}
\end{equation}
Eq. (\ref{integral1}) can be transformed into
\begin{equation}
\sigma_{xy}(0,\tau_z)=\tau _{z}\frac{e^{2}}{2h}\int_{0}^{\infty }d(k^{2})%
\frac{\partial (\cos \theta )}{\partial (k^{2})}
\end{equation}
The values of $\cos \theta $ at $k=0$ and $k\rightarrow \infty $ only depend
on the signs of $B$ and $\Delta $, respectively. As a result, in the
insulating regime $-\frac{|\Delta |}{2}\leq \mu \leq \frac{|\Delta |}{2}$,
we find that the anomalous Hall conductance for each hyperbola has the form
\begin{equation}  \label{chernnumber}
\sigma _{xy}(0,\tau_z)=-\tau_z\frac{e^{2}}{2h}(\mathrm{sgn}(\Delta )+\mathrm{%
sgn}(B)),
\end{equation}%
where $\frac{\tau_z}{2}(\mathrm{sgn}(\Delta )+\mathrm{sgn}(B))$ is the Chern
number of the valence bands.\cite{Thouless82} It is a well known result in
the field theory that the Hall conductance of the massive Dirac fermions is
a half of quanta $e^{2}/h$, $\sigma _{xy}=-\tau_z\frac{e^{2}}{2h}\mathrm{sgn}%
(\Delta )$ (if $B=0$). (see Ref. \onlinecite{Redlich-84prd}.) Our result
demonstrates that a non-zero quadratic term in $k$ ($B\neq 0$) will give a
reasonable result since a half quantized Hall conductance is not possible
for a non-interacting system. For small $L$, the parameter $B$ is always
negative, and $\Delta $ changes its sign accompanying a gap close-and-reopen
while the thickness of the thin film increases [Fig. \ref{fig:gap}(f)]. The
sign change of $\Delta $ flips the $\mathbf{k}$-dependent spin configuration
near the $\Gamma $ point and results in a jump $\tau_z$ in the Chern number,%
\cite{Shen-04prb,Zhou-07epl} \emph{i.e.}, a topological quantum phase
transition as discussed in quantum spin Hall effect in HgTe/CdTe quantum
well.\cite{Bernevig-06Science} In the ultrathin limit, $\Delta >0$ and the
Hall conductance is zero, while $\Delta <0$ at a larger thickness and the
Hall conductance $\sigma _{xy}(0,\tau_z)=\tau_z(e^{2}/h)$, as shown in Fig. %
\ref{fig:hall conductance}. By the edge-bulk correspondence,\cite{Hatsugai93}
gapless helical edge-states shall appear accompanying such a transition.
This result is indeed supported by the solutions of the differential
equation of $h_{\pm }(k)$ in a geometry of a semi-infinite plane, in which
there exists a gapless edge state only when $\Delta /B>0$, and the sign of
the Hall conductance can be justified from the chirality of the edge states.

When the Fermi level $\mu $ lies in the electron band, according to Eq. (\ref%
{hallformula})
\begin{equation*}
\sigma _{xy}(\mu >\frac{|\Delta |}{2},\tau _{z})=\sigma _{xy}(0,\tau
_{z})+\tau _{z}\sigma _{xy}^{+}
\end{equation*}%
where at zero temperature,
\begin{equation*}
\sigma _{xy}^{+}=\frac{e^{2}}{4h}\int_{0}^{k_{F}}d(k^{2})\frac{A^{2}(\frac{%
\Delta }{2}+Bk^{2})}{[A^{2}k^{2}+(\frac{\Delta }{2}-Bk^{2})^{2}]^{3/2}}.
\end{equation*}%
Near the $\Gamma $ point, the $B$ term in the electron band dispersion can
be neglected, so that $\varepsilon _{c}^{2}\approx (\hbar v_{F})k^{2}+(\frac{%
\Delta }{2})^{2}$, and
\begin{equation*}
k_{F}^{2}=[\mu ^{2}-(\frac{\Delta }{2})^{2}]/(\hbar v_{F})^{2}.
\end{equation*}%
After a straightforward calculation and Taylor expansion of the result up to the linear term of $\mu$, the Hall conductance is obtained when
the Fermi surface is in the conduction bands,
\begin{equation}
\sigma _{xy}(\mu >\frac{|\Delta |}{2},\tau _{z})=-\tau _{z}\frac{e^{2}}{2h}[%
\mathrm{sgn}(B)+\mathrm{sgn}(\Delta )-\frac{8\pi (\hbar v_{F})^{2}\delta n}{%
\Delta |\Delta |}].  \notag
\end{equation}%
where the Fermi level $\mu$ is replaced by $\delta n$, the carrier density near the band bottom, by using 
the relation $|\Delta|(\mu-\frac{|\Delta|}{2})/(4\pi \hbar^2 v_F^2) \approx \delta n$, which is found by using the dispersion $\varepsilon_c$ and the same Taylor expansion approach.
One can check
that when the Fermi level lies in the valence bands, $\sigma _{xy}$ is the
same as the above result due to the particle-hole symmetry.

As shown in Fig. \ref{fig:gap}(e), $B$ is negative for small $L$. For $%
\Delta >0$ where the intrinsic film is a trivial insulator, doping with
electrons or holes can result in a metallic spin Hall effect, with the Hall
conductance given by
\begin{eqnarray}
\sigma _{xy}(\mu>\frac{\Delta}{2},\tau_z)=\tau_z\frac{e^{2}}{h}\frac{4\pi }{%
\Delta ^{2}}(\hbar v_{F})^{2}\delta n,
\end{eqnarray}
where $\delta n$ is the carrier density. 
In the two Dirac hyperbolas, the
Hall conductance are opposite (see Fig. \ref{fig:hall conductance}) and spin
vectors near the band bottom point in the $+z$ and $-z$ directions
respectively. Thus net spin accumulations with out-of-plane polarization are
expected on the two edges of the thin film. For $\Delta <0$, doping reduces
the quantized Hall conductance and we find
\begin{eqnarray}
\sigma _{xy}(\mu>\frac{|\Delta|}{2},\tau_z)=\tau_z\frac{e^{2}}{h}[ 1-\frac{%
4\pi }{\Delta ^{2}}(\hbar v_{F})^{2}\delta n].
\end{eqnarray}
The Hall conductance as a function of the Fermi level $\mu $ are plotted in
Fig. \ref{fig:hall conductance}. Note that because the Hall conductances for the two hyperbolas are always equal in magnitude and opposite in sign, and the two hyperbolas have right opposite spin orientations, here the Hall conductances are referred to as the ordinary and quantum spin Hall effects.

\section{\label{sec:optical}Spin optical selection rule}

Since an energy gap is opened in the Dirac hyperbolas, interband transitions
between the conduction and valence surface bands can be excited by optical
field. In the two gapped Dirac hyperbolas being time-reversal of each other,
interband transition couples preferentially to right-handed ($\sigma +$) or
left-handed ($\sigma -$) circular polarized light, as was first discovered
in the context of graphene.\cite{Yao-08prb} In the thin film topological
insulator where the two Dirac hyperbolas are associated with opposite spin
configurations (Fig. \ref{fig:meron}), such an optical transition selection
rule is of significance since it allows the spin dynamics to be addressed by
optical means. The interband couplings in the two hyperbolas to normally
incident circular polarized lights can be studied by calculating the degree
of circular polarization, defined as
\begin{equation*}
\eta (\mathbf{k})\equiv \frac{|\pi _{cv}^{+}(\mathbf{k})|^{2}-|\pi _{cv}^{-}(%
\mathbf{k})|^{2}}{|\pi _{cv}^{+}(\mathbf{k})|^{2}+|\pi _{cv}^{-}(\mathbf{k}%
)|^{2}},
\end{equation*}%
where $\pm $ corresponding to $\sigma \pm $ lights, respectively, and the
interband matrix element of the velocity operator is defined by
\begin{equation*}
\pi _{cv}^{\pm }(\mathbf{k})\equiv \langle u_{c}(\mathbf{k})|\frac{\partial
h_{\tau _{z}}}{{\partial k_{x}}}\pm i\frac{\partial h_{\tau _{z}}}{{\partial
k_{y}}}|u_{v}(\mathbf{k})\rangle ,
\end{equation*}%
where $u_{c}(\mathbf{k})$ and $u_{v}(\mathbf{k})$ are the eigen states for
the conduction and valence bands of $h_{\tau _{z}}$, and have already been
given in Eq. (\ref{eigenstates}). By ignoring $Bk^{2}$ terms near the $%
\Gamma $ point, one finds
\begin{equation*}
|\pi _{cv}^{\pm }(k)|^{2}\simeq (\hbar v_{F})^{2}(1\pm \tau _{z}\cos \theta
)^{2},
\end{equation*}%
where $\cos \theta =\Delta /(\varepsilon _{c}(k)-\varepsilon _{v}(k))$, with
$\varepsilon _{c}$ ($\varepsilon _{v}$) the dispersion of the conduction
(valence) band of $h_{\tau _{z}}$ given in Eq. (\ref{eigenenergies}). Then
the degree of polarization is found out for each $\tau _{z}$
\begin{equation*}
\eta (k)=\tau _{z}\frac{2\cos \theta }{1+\cos ^{2}\theta }.
\end{equation*}%
Near the Dirac hyperbola center where $\varepsilon _{c}(k)-\varepsilon
_{v}(k)\simeq |\Delta |$, $\sigma +$ ($\sigma -$) light couples only to the
Dirac hyperbola $\tau _{z}=$ sgn$(\Delta )$ ($\tau _{z}=-$sgn$(\Delta )$),
causing transition between the spin down (up) valence state and the spin up
(down) conduction state. This is in sharp contrast to the spin optical
transition selection rule in conventional semiconductor which leaves the
spin part of the wavefunction unchanged. This spin optical transition
selection rule makes possible optical injection of spins in the thin film.
For example, with band-edge optical excitation by $\sigma +$ circularly
polarized light, spin up electrons in the conduction band and spin up holes
(by convention, a spin up hole refers to an empty spin down valence state)
in the valence band are created in the Dirac hyperbola $\tau _{z}=\mathrm{sgn%
}(\Delta )$, which can be separated by an in-plane electric field to prevent
the radiative recombination.

\section{\label{sec:conclusion}Conclusions}

In this paper, we derive an effective model for an ultrathin film of
topological insulator, in which tunneling between the surface states on the
top and bottom surfaces opens a finite gap in the Dirac cone centered at the
$\Gamma $ point ($k=0$). The low lying physics of the ultrathin film can be
described as two degenerate massive Dirac hyperbolas which form time
reversal copy of each other. Each massive band has a $\mathbf{k}$-dependent
spin configuration: one near the $\Gamma $ point determined by the energy
gap, and the other at $k$ large enough determined by a spin-orbit coupling
term quadratic in $k$. It is found that the energy gap oscillates with the
thin film thickness, and changes sign at critical thicknesses. Across the
transition points, the $\mathbf{k}$-dependent spin configuration near the $%
\Gamma $ point is flipped while those at large $k $ remains unchanged,
leading to a topological quantum phase transition where the Chern numbers of
the surface bands change [Eq. (\ref{chernnumber})]. In the two Dirac
hyperbolas of opposite spin configurations, the $\mathbf{k}$-dependent spin
structure results in a distribution of orbital magnetic moment and Berry
curvature with opposite signs. In doped metallic regime, the Berry curvature
drives the spin Hall effect of the extra carriers which leads to net spin
accumulations on the thin film edges. We also discover a spin-dependent
optical transition selection rule which allows optical injection of spins by
circular polarized lights into the thin film.

This work was supported by the Research Grant Council of Hong Kong under
Grant No.: HKU 7037/08P, and HKU 10/CRF/08.

\emph{Note added}: After posting this paper on arXiv, we learnt about the
works by Liu \emph{et al}\cite{Liu2009.arxiv} and Linder \emph{et al}\cite%
{Linder2009.arxiv}, in which the similar finite size effect of the surface
states was studied.


\end{document}